\documentclass[a4paper,10pt]{article}
\usepackage{color,cite,graphicx}
\usepackage{setspace}
\usepackage{latexsym,amssymb,epsf}
\newcommand{\be}{\begin{equation}}
\newcommand{\ee}{\end{equation}}
\newcommand{\ba}{\begin{eqnarray}}
\newcommand{\ea}{\end{eqnarray}}
\newcommand{\nn}{\nonumber}
\newcommand{\qq}{\qquad}
\newcommand{\lb}{\label}

\newcommand{\hf}{{1\over 2}}
\newcommand{\dd}{{\rm d}}
\newcommand{\eq}{\equiv}
\newcommand{\app}{\approx}
\renewcommand{\d}{\partial}

\newcommand{\ff}{\Phi}
\newcommand{\rhoh}{\rho}

\renewcommand{\r}{{\bf r}}
\newcommand{\ri}{\r_i}
\newcommand{\rj}{\r_j}
\newcommand{\rij}{r_{ij}}
\newcommand{\rrij}{{\bf r}_{ij}}
\newcommand{\rp}{|\r-\r'|}
\newcommand{\drr}{\dd\r\dd\r'}
\renewcommand{\i}{{\bf i}}
\renewcommand{\j}{{\bf j}}
\renewcommand{\l}{{\bf l}}
\newcommand{\ffi}{\ff_\i}
\newcommand{\fii}{{\bf F}_i}
\newcommand{\q}{{\bf q}}

\newcommand{\erf}{{\rm erf}\,}
\newcommand{\erfc}{{\rm erfc}\,}
\newcommand{\xcut}{x_{\rm cut}}
\newcommand{\rcut}{r_{\rm cut}}
\newcommand{\short}{{\rm short}}
\newcommand{\self}{{\rm self}}
\newcommand{\lng}{{\rm long}}

\newcommand{\nshort}{N_{\rm short}}

\newcommand{\kopf}{{\rm HEAD}}
\newcommand{\tail}{{\rm TAIL}}

\newcommand{\mmax}{{\rm MMAX}}

\newcommand{\allreduce}{{\rm MPI\_ALLREDUCE}}

\begin{document}

\begin{center}


{\huge Particle-Particle, Particle-Scaling function (P3S) algorithm for electrostatic problems in free boundary conditions.}

\vspace{1cm}
Alexey Neelov, S. Alireza Ghasemi, Stefan Goedecker 

{\small\it Institute of Physics, University of Basel, Klingelbergstrasse 82, CH-4056 Basel, Switzerland}

\begin{abstract}
An algorithm for fast calculation of the Coulombic forces and energies of point particles with free boundary conditions is proposed. Its calculation time scales as $N \log N$ for $N$ particles. This novel method has lower crossover point with the full $O(N^2)$ direct summation than the Fast Multipole Method. The forces obtained by our algorithm are
analytical derivatives of the energy which guarantees energy conservation during a molecular dynamics simulation.
Our algorithm is very simple. An MPI parallelised version of the code can be downloaded under the GNU General Public License from the website of our group.

\end{abstract}
\end{center}
\section{Introduction}

The computation of the electrostatic or gravitational interaction of a large number $N$ of point particles is a central problem in many fields of physics, such as molecular dynamics \cite{md} and astrophysics \cite{spurzem}.

If the charge distribution is nonperiodic, one frequently employs the fast multipole method (FMM)  \cite{rokhlin}-\cite{pfalzner}. Its computation time scales as $O(N)$. Its point of the crossover with
full direct calculation depends on the accuracy and is estimated to be $10^3-10^4$ in \cite{Headgordon} or 
$10^4-10^5$ in \cite{esselink}. The Barnes and Hut hierarchical tree method \cite{barnes} is advocated for the use with low accuracy ($0.1-1\% $) in \cite{gibbon}.

 The cutoff methods were recently revived \cite{wolf},
\cite{fennell}. They also scale as $O(N)$, but because of the immense prefactor they are only used for special systems. 

If the charge distribution is periodic, one can use the Ewald method \cite{ewald}-\cite{board}, that scales as $O(N^{3/2})$. It is good for up to $10^3-10^4$ particles and can achieve high accuracy; it is clearly faster than
full direct summation of the periodic particle images.

The Ewald method was further improved by the use of Fast Fourier Transform; that led to the particle-mesh schemes \cite{hockney}-\cite{split}. These algorithms scale as $O(N \log N)$.
The same scaling was obtained by combining the Ewald method with the FMM for the calculation of the short range forces \cite{duan} and, alternatively, by using nonuniform FFT in the Ewald method \cite{hedman}. 
The particle-mesh schemes are suitable \cite{gibbon} for relatively high RMS force error of about $10^{-4}$ and $N=10^4-10^5$.

The FMM method can be applied in the periodic case too \cite{shimada},\cite{pollock}, 
\cite{schmidt} -\cite{challacombe}
  especially if one needs high precision.
Due to its better scaling, it becomes preferable in the limit of large $N$. However for not so big $N$ the particle-mesh schemes are faster. The exact location of the crossover point depends both on the system studied and the computer used \cite{gibbon}, \cite{challacombe}. The usual estimate for the crossover is around $N=10^5-10^6$.

It is mentioned in \cite{md},\cite{schulten1} that in the FMM the forces are not equal to the negative analytical gradients of the energy. Therefore if one uses the FMM for an MD simulation, the total energy will change during the
simulation, unless very high precision is used - see e.g. Fig.1 of \cite{figuerido}. Thus it is impossible to do simulations in the microcanonical ensemble.

In the particle-mesh schemes it is possible to have conservation of either energy or momentum, but not the two
together \cite{spme}, \cite{pm}. The particle-mesh schemes are usually also easier to code, compared to the FMM.

Much attention has been given to the parallelization of the particle-mesh \cite{pollock},\cite{sagui}, \cite{touk}-\cite{dubna}, and especially FMM
\cite{schulten92}-\cite{ogata} algorithms. In the case of FMM, the parallelization is more difficult technically, but more efficient. Recent papers \cite{kurzak1}, \cite{kurzak2} cite speedups of over 100 for $10^5$ particles for a state of the art parallel implementation of FMM. In contrast, the parallel  speedups of the simple particle-mesh methods of \cite{touk},\cite{dubna} do not exceed 10 for $10^5$ particles; there is probably room for improvement. 

Many systems ranging from those in the electronic structure calculations of molecules to those of astrophysics require the use of free boundary conditions (BC). Therefore the imposition of artificial periodicity that is needed for the Ewald-type methods can lead to complications \cite{art1}, \cite{art2}. It would be useful to have a particle-mesh-like method that could handle free BC.

Such algorithms were set forth in \cite{hockney70}, \cite{martyna}, \cite{sutmann}. The Hockney
method \cite{hockney} was intended for applications in plasma physics where high precision is not required. 
Martyna and Tuckerman \cite{martyna} used the SPME method of Essman et al. \cite{spme}, along with a special scheme to handle free BC. That scheme was also used for the solution of Poisson equation for smooth charge distributions occurring in electronic structure calculations.
However, the method of \cite{martyna} has the disadvantage that the accuracy decreases if one approaches the border of the cell. Thus one needs to take bigger cells that leads to an increase of the CPU time needed.

Another line of the  particle-mesh like algorithms stems from the Fast Fourier Poisson method \cite{ffp},\cite{sagui}. In that case one solves the Poisson equation for
the auxiliary charge distribution that is made up from Gaussians centered at the
original particle positions. The resulting charge density is then projected onto a grid and the thus discretized Poisson equation is solved via an FFT. The forces
are obtained by analytical differentiation of the energy expression; this solves
the aforementioned problem of energy conservation. This method is believed to be accurate but slow \cite{md} compared to the standard particle-mesh schemes such as P3M \cite{hockney} or SPME \cite{spme}.

It was proposed in \cite{beckers} to use multigrid \cite{multigrid} for the solution of the Poisson equation, instead of the Fast Fourier Transform. This technique was further developed in \cite{sagui},\cite{banerjee},\cite{split}. The CPU time in the multigrid algorithm scales as $O(N)$; however
due to large prefactor it becomes preferable to the FFT-based methods only for a very large $N$ and/or on
a massively parallel computer.

The above Fast Fourier-Poisson methods are only applicable with the periodic BC. However, in the new method of Sutmann and Steffen \cite{sutmann} the free BC were accounted for by calculating the
potential at the boundaries with the FMM algorithm. This algorithm may be a competitor to the FMM algorithm in the future. Another particle-mesh-like algorithm
that uses an iterative solver instead of FFT is described in \cite{skeel}.

In a recent paper \cite{psolver} we proposed a new algorithm for the solution of the Poisson equation for smooth charge distributions with free BC.  It is also MPI-parallelised.

In the present paper we propose a combination of the Poisson solver with free BC \cite{psolver} with the Fast Fourier Poisson method \cite{ffp}. We call it particle-particle particle-scaling function (P3S) algorithm to emphasize the relation to the particle-mesh schemes. Our method  can achieve high accuracy and speed and compete with the FMM schemes in the range of particle numbers $N=10^3-10^5$ that is important in many applications.
The approximate forces resulting from our method are exact (negative) analytic gradients of the approximate energy, which allows the conservation of energy during  MD runs.

To calculate the short-range forces, we make a linked list \cite{allen}. Then, following  \cite{heinz},\cite{yao}, we rearrange the particles to optimize the cache performance.

An MPI parallelised version of our code can be downloaded under the GNU General Public License from the website of our group,\\ http://pages.unibas.ch/comphys/comphys/SOFTWARE/index.html.

The paper is organized as follows. Section 2 describes the Ewald construction in
free BC. In Section 3 we briefly review the Poisson solver of \cite{psolver}
for the calculation of the long range Ewald forces and energies for free BC . In Sections 4 and 5 we discuss the cutoffs for the short and
long range forces, respectively. In Section 6 we describe the linked cell list for the
acceleration of the short range force calculation. Section 7 contains the final formulas 
for the interparticle forces. In  Section 8 we give the results of our code for systems
of 1000-20000 particles and give the graphs of the optimal parameter values for a given accuracy. Section 9 describes the results of an MD simulation of a 1000-particle NaCl crystal
that demonstrates the energy conservation property of our algorithm. Finally, in Section 10
we describe an efficient parallelization of our algorithm and present the parallel speedup results on a CRAY XT3.

\section{The Ewald construction in free boundary conditions \cite{pollock}.}

The total electrostatic energy of $N$ point charges in free BC is given by
\ba
U={1\over 2} \sum_{i=1}^N\sum_{j\ne i}^N {Q_i Q_j\over \rij},\qq
\rij\eq|\ri-\rj|. \nn
\ea

Adding and subtracting the term corresponding to the electrostatic energy of smooth point charges with density $\rhoh_i(\r)$, we get:

\ba
&&U={1\over 2} \sum_{i=1}^N\sum_{j\ne i}^N\biggl[{Q_i Q_j\over \rij}-
\int {\rhoh_i(\r)\rhoh_j(\r')\over |\r-\r'|}\drr \biggr]+\nn\\
&&+
{1\over 2} \sum_{i=1}^N \sum_{j=1}^N\int {\rhoh_i(\r)\rhoh_j(\r')\over |\r-\r'|}\drr-
\nn{1\over 2} \sum_{i=1}^N \int {\rhoh_i(\r)\rhoh_i(\r')\over |\r-\r'|}\drr.\lb{u}
\ea
The Ewald choice  for the screening charge distribution is
\ba
\rhoh_i(\r)=\gamma(\r-\r_i),\qq
\gamma(\r)\eq Q_i (G^2/ \pi)^{3/2}\exp [-G^2\r^2].\lb{rhoi}
\ea
We have also experimented with the variant 
$\gamma(\r)=A (\rcut^2-r^2)^m$; $m=4,8,16$, where the factor $A$ normalizes the charge to one. However, the CPU time spent by the program with this screening distribution was slightly longer than with the Gaussian for the same accuracy.

The sum of the screening charge distributions is 
 \ba
 \rhoh(\r)=\sum_{i=1}^N \rhoh_i(\r).\lb{rho}
 \ea
Then, (\ref{u}) assumes the form
\ba
U&=&E_{\short}+E_{\lng}-E_{\self},\lb{uee}\\
E_{\short}&=&{1\over 2} \sum_{i=1}^N\sum_{j\ne i}^N {Q_i Q_j\over \rij}\erfc \biggl({G\rij\over \sqrt{2}}\biggr),\lb{eshort}\\
E_{\lng}&=&\hf \int {\rhoh(\r)\rhoh(\r')\over |\r-\r'|}\drr,\qq
E_{\self}={G\over\sqrt{2\pi}}\sum_{i=1}^N Q_i^2.\lb{el}
\ea

\section{Calculation of the long range energy using the interpolating scaling functions.}

The long range term $E_{\rm long}$ in (\ref{el}) is nothing but the electrostatic energy
of the smooth charge distribution $\rhoh(\r)$. Therefore it can be evaluated by a suitable Poisson solver. We make use of the one from Ref. \cite{psolver} with free BC.
It amounts to expanding the screening charge density (\ref{rho}) in a real space basis defined on the grid with spacing $h$:
\ba
\rhoh(\r)&\app&\widetilde\rho(\r)=\sum_{\i}\rho_\i\ffi(\r),\qq \i\eq (i_1,i_2,i_3),\lb{rrho}\\
\ffi(\r)&=&\ff(x/h-i_1)\ff(y/h-i_2)\ff(z/h-i_3).\nn
\ea
  It was suggested in \cite{psolver} to take the interpolating scaling functions  \cite{DD},\cite{Gbook} of high order (up to 100) as the basis functions $\phi(x)$. The scaling functions of the order $N$ interpolate the polynomials
of the order $N$ exactly and are reasonably localized. Therefore they can interpolate a Gaussian very well.

On the other hand, since the scaling functions are cardinal, we obtain for the coefficient in (\ref{rrho}):
\ba
\rho_\i=\rhoh(\i h).\lb{rhoii}
\ea
The action of calculating this screening distribution on the grid is called
the charge assignment in the standard P3M schemes.

Consider the potential that arises from the approximate charge distribution $\widetilde\rho(\r)$ in (\ref{rrho}):
\ba
\phi(\r)=\int {\widetilde\rho(\r)\over \rp}\dd\r'. \lb{phiappr}
\ea

At a grid point $\j$, this potential has the form
\ba
\phi(\j h)\eq\phi_\j=h^2\sum_{\i} K_{\i-\j} \rho_\i,\nn
\ea
where the kernel 
\ba
K_\i\eq\int {\ffi(\r)\dd\r\over |\r|}\nn
\ea
is computed in \cite{psolver}.

From this moment on we can use the grid sum approximation to the long range energy:
\ba
E_{\rm long}\app {h^3\over 2}\sum_\i \rho_\i \phi_\i= {h^5\over 2}\sum_{\i\j} \rho_\i\rho_\j K_{\i-\j}.\lb{rkr}
\ea
The latter sum is a convolution that can be calculated via FFT techniques \cite{psolver}. The kernel $K_\i$ is calculated only once at the beginning of a calculation
and does not change. Thus the use of high order of interpolation does not lead to a significant increase of calculation time \cite{psolver}. 

\section{The long-range part cutoff.}

The density array in (\ref{rkr}) is defined via (\ref{rhoii}):
\ba
\rho_\i=\rhoh(\i h)=\sum_{j=1}^N\gamma(\i h -\rj).\nn
\ea
However, the Gaussian (\ref{rhoi}) is a quickly decaying function. Therefore, one can
make the summation in the above equation only for the charges within the distance
$\xcut$ from the grid point:
\ba
\rho_\i\app\sum_{|\i h -\rj|<\xcut}\gamma(\i h -\rj).\lb{rrhoi}
\ea
The cutoff distance is chosen so that the function (\ref{rhoi}) has a small value at the radius $\xcut$.

Actually, the sum in (\ref{rrhoi}) is calculated slightly differently in our program. The charge position $\rj$ is replaced by the position $h {\bf i}_j$ of the grid point
that is nearest to it:
\ba
\rho_\i\app\sum_{(\i -\i_j)^2\,<\,\xcut^2/h^2}\gamma\bigl(h(\i -\i_j)\bigr).\lb{rhoapp}
\ea
The error entailed by this is negligible. Then we precalculate and store the square roots of integers in order to determine to which gridpoints a given charge contributes. 
This is faster than calculating square roots of real numbers and truncating
to integers in (\ref{rrhoi}).

As a result the calculation time of the charge spreading and long-range force interpolation is reduced allmost by a factor of two compared to a simple summation over a rectangular domain. Note that this is only possible because in the FFP algorithm the charge density assigned to the grid
is made of Gaussians. In other particle-mesh schemes B-splines are used instead,
so our method would be inapplicable there.

Making the cutoff approximation in (\ref{rkr}) gives us, finally:
\ba
E_{\rm long}\app {h^5\over 2}\sum_{\i\j} \rho_\i\rho_\j K_{\i-\j}.\lb{elong}
\ea

The long-range energy of our algorithm is given only by the
formulas (\ref{rhoapp}),(\ref{elong}). Our algorithm is thus as simple in formulation as the Fast Fourier Poisson method \cite{ffp}.
On the other hand, the properties of interpolating scaling functions allow
us to combine fast Gaussian charge assignment and very high order interpolation. This is to be contrasted e.g. to the use of Lagrange interpolation in the PME method \cite{pme} where the values  of the interpolating functions have to be actually computed at the particle positions. A good account of this method is
given e.g. in Ref. \cite{petersen}. If one uses Lagrange interpolation
of the order $L$, then the length of the Lagrange interpolation filter is $L+1$ 
and the number of grid points to which a single charge is assigned is thus $(L+1)^3$. For
large $L$ this means that a lot of CPU time will be spent on the charge
assignment and force back interpolation. To avoid this, in the PME one uses only low order interpolation.

However, we would like to repeat that our method treats systems with free BC 
whereas the standard particle-mesh ones are primarily for the periodic systems.

To illustrate the accuracy of our approximation, we calculate the
self-energy of a single unit Gaussian (\ref{rhoi}) by the formula (\ref{elong}). The Gaussian is initially at a grid point and then gradually shifted by two grid constants to see the dependence of the error on its position relative to the grid. We have $\xcut=4.5h$ and $h=0.7$; these are typical parameter values that we use in other tests .
 
\begin{figure}[htp]
\centering
\includegraphics[totalheight=.4\textheight,]{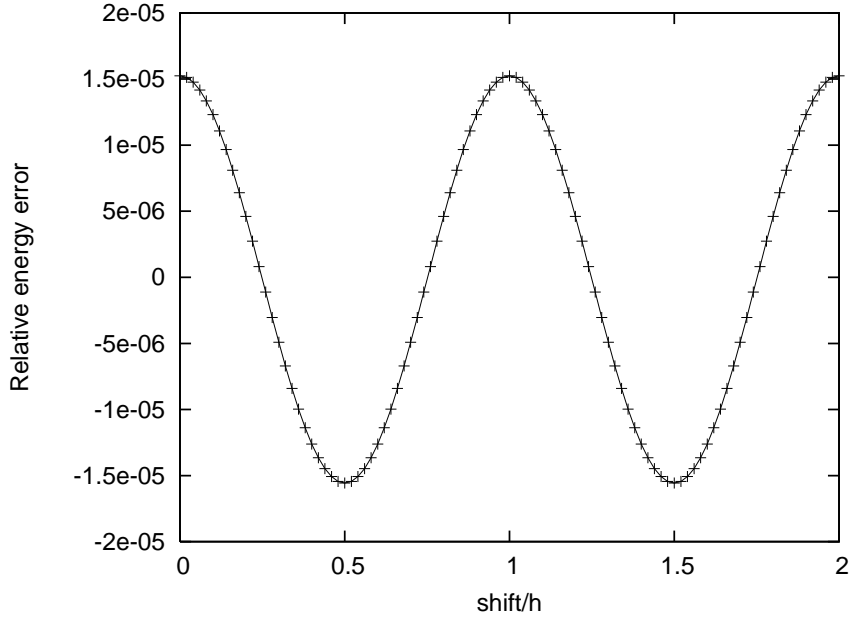}
\caption{Relative error in the self-interaction energy of a single Gaussian .}\label{fig:ener}
\end{figure}

 The graph of the resulting energy error vs. the shift distance
is given at Fig. \ref{fig:ener}.
We see that the error is rather smooth; there is no visible jumps coming from the cutoff approximation. The energy oscillation is due to the grid discretization;
its amplitude goes to zero in the limit $h\to 0$

\section{The short-range part cutoff.}

The short-range pairwise potential in (\ref{eshort}) is a rapidly decaying
function. Therefore, it is standard to introduce the cutoff radius $\rcut$ beyond which
the short-range potential is approximated by zero:
\ba
E_{\short}&=&{1\over 2} \sum_{i,j:\,\rij<\rcut} {Q_i Q_j\over \rij}\erfc \biggl({G\rij\over \sqrt{2}}\biggr).\lb{eshort1}
\ea
To summarize, the Coulombic energy of $N$ charges in our approximation has
the form (\ref{uee}) with the approximate long and short contributions given
by (\ref{elong}),(\ref{eshort1}), correspondingly.

The evaluation of the error function in (\ref{eshort1}) is
computationally inefficient, compared to algebraic operations. Therefore, we
replace the term that corresponds to error function by a spline approximation in the interval $(0,\rcut)$.

The function $\erfc (G r/\sqrt{2})/r$ has a singularity at 0. Therefore we rewrite it as
\ba
\erfc (G r/ \sqrt{2})/r=1/r-\erf(G r/ \sqrt{2})/r,\lb{erfc}
\ea
and approximate with a spline only the second term which is regular. This approximation holds for $r<\rcut$. For $r>\rcut$, the whole potential $\erfc (G r/ \sqrt{2})/r$ is set to 0.

\section{Calculating the forces.}

The force acting on the $i$-th particle is given by the gradient of (\ref{uee}):

\ba
\fii={\d U\over\d \ri}=\fii^{\short}+\fii^{\lng};\hspace{5.8cm}\nn\\
\fii^{\short}=Q_i{\sum_{j:\rij<\rcut}}'Q_j\biggl({G\over \rij^2}\sqrt{2\over\pi}\exp \biggl[-{G^2\rij^2\over 2}\biggr]+{1\over\rij^3}\erfc\biggl[{G\rij\over\sqrt{2}}\biggr]
\biggr)\rrij,\lb{fshort}
\ea 
where $\rrij=\ri-\rj$; the prime at the sum indicates that it is performed
over $j\ne i$.
The short range forces (\ref{fshort}) are equal to the negative analytical
gradients of the short range energy (\ref{eshort}). They are also calculated via a spline approximation. Actually, the spline that we use for (\ref{erfc}) is obtained by integration of the spline for (\ref{fshort}).

Following \cite{sagui}, we also get the long range forces as the negative analytical gradients of (\ref{elong}):
\ba
\fii^{\lng}\app h^5\sum_{\l\j} {\d\rho_\l\over\d \ri}\rho_\j K_{\l-\j}\eq h^5\sum_{\l\j} \q (\l h -\ri)\rho_\j K_{\l-\j},\lb{flong}
\ea
where
\ba
\q(\r)={\d\gamma(\r)\over\d \r}=-2\r G^2\gamma(\r)\nn
\ea
is the vector of derivative Gaussians. The calculation of the long range forces (\ref{flong}) is called the force back interpolation.

Taking into account the cutoff (\ref{rhoapp}), one can rewrite the long range force as
\ba
\fii^{\lng}= h^5\sum_{\j,\l:(\l -\i_j)^2\,<\,\xcut^2/h^2} \q (\l h -\ri)\rho_\j K_{\l-\j}.\nn
\ea

\section{Linked cell list.}

In order to avoid the scanning of all particle pairs in (\ref{eshort1}) we make a linked subcell list \cite{allen}. The subcell size is smaller or equal
to $\rcut/\mmax$ where $\mmax$ is a positive integer ranging, in practice, from 1 to 3. The standard linked cell list algorithm corresponds to $\mmax=1$; higher values of $\mmax$ correspond to smaller cells such as those of \cite{heinz},\cite{yao}. To account for the free BC we add $\mmax$ layers of empty cells outside the original cell. The sum of forces for a given particle is then done over all particles in
subcells which are within $\rcut$ of the subcell where the original
particle is located.

The sum over subcells is done along the stripes in the direction $x$. 
In particular, following  \cite{heinz},\cite{yao}, we rearrange the particles so that the particles in the same subcell have consecutive numbers and the subcells are sorted in the $x$ direction. The actual sorting can be avoided
because we already have the linked subcell list. In more detail, 

\begin{itemize}
\item The first and last particle numbers in each subcell are stored in separate arrays $\kopf$ and $\tail$.

\item If the cell is empty, its $\kopf$ value is that of the neighboring cell in the direction of increasing $x$ coordinate. For the cells outside the simulation box (to the right of it), $\kopf=N+1$ by default. 

\item The $\tail$ value of an empty cell is that of its neighbor in the direction
of decreasing $x$. For the outside cells (to the left of the simulation box), $\tail=0$.

\item The loop over all particles in the cell stripe:
\ba
(i_1,i_2,i_3),(i_1+1,i_2,i_3),\dots,(j_1,i_2,i_3)\nn
\ea
is done as a loop the over particle numbers from $\kopf(i_1,i_2,i_3)$ to $\tail(j_1,i_2,i_3)$. In this way it does not matter whether empty cells occur.
\end{itemize}

The reordering of the particles has the additional advantage that
the cache performance of the long range part is optimized. The reason is that the particles that have close indices are also close physically: $|\r_i-\r_j|$ is small when $|i-j|$ is small.

The loop in the charge assignment and force back interpolation goes over the particle numbers. The particles in the same subcell have overlapping screening
charge densities, so the elements of the density/potential array are reused again and again, as the loop index goes over the particles in the same subcell.

This makes the program significantly faster compared to the case without
reordering, especially if the particles are distributed randomly.

\section{The results of the serial code.}
We chose two test systems:
\begin{itemize}
\item $N$ particles in the unit cube with random coordinates and charges equal to $\pm 1$, the total charge being zero (henceforth called the random system)
\item $N$ particles with charges $\pm 1$ forming a rock salt crystal, filling
the unit cube with $M\app N^{1/3}$ nodes at each side. The lattice constant is then  $d=(M-1)^{-1}$. Each particle is then shifted
away from its node by a vector with random coordinates in the range $(-d/3,d/3)$. This system will be called the crystal one.
\end{itemize}
In both cases, the particle number $N=1000*10^{j/3}$; $j=0\dots 4$. In other words, we consider the following values of $N$: $1000,2154,4642,10000,21544$.

We always use the scaling functions of the order 100. We found that using lower orders
leads to a decline in accuracy, while the order of
the scaling functions only affects the time of the calculation of the kernel which is done only once for a simulation.

The values of $\xcut$, $\rcut$, $G$ and $h$ are free parameters. As explained e.g. in \cite{petersen}, such parameters should be chosen such that
\begin{itemize}
\item The accuracy is fixed to some value.
\item The CPU time spent is minimal for the given accuracy.
\end{itemize}

As the  measure of the accuracy we use the mean square force error:
\ba
\delta F\eq\sqrt{\sum_{i=1}^N ({\bf F}_i-{\bf F}_i^{\rm direct})^2\over
\sum_{i=1}^N ({\bf F}_i^{\rm direct})^2},\nn
\ea
where 
\ba
{\bf F}_i^{\rm direct}=Q_i\sum_{j\ne i}Q_j{\r_i-\r_j\over|\r_i-\r_j|}\nn
\ea 
are the forces obtained by the full direct calculation.

Unfortunately we cannot give the {\it a priori} error estimates of the accuracy. The error in the short-range forces is the same as that for the Ewald method, and is estimated in \cite{kolafa},\cite{petersen}. However, our calculation of the long range forces involves the 
use of a finite basis and the approximation in Eq. (\ref{rkr}), the accuracy of which is difficult to estimate.

Therefore, for our two test systems we made tests with all reasonable values of the parameters. The most important parameter is $G$, the width of the Gaussian in (\ref{rhoi}). We also introduced the following dimensionless quantities: the dimensionless Gaussian cutoff $G\xcut$; the cutoff ratio  $\rcut/\xcut$, the dimensionless grid constant $G h$ and
\ba
N_{\rm short}\eq{ 4/3\pi \rcut^3\over V_{\rm box}} N.\nn
\ea
The latter is the average number of particles inside a sphere of radius $\rcut$. It turns out that the relative MSQ force error depends stronger on these dimensionless quantities and $G$ than on the number of particles in the system.

We checked 7-10 possible values for each parameter. From the resulting pool of results we selected the so called Pareto frontier. A point is on the Pareto frontier if there is no other point
which has both smaller CPU time and smaller error.

\begin{figure}[htp]
\centering
\includegraphics[totalheight=.4\textheight,]{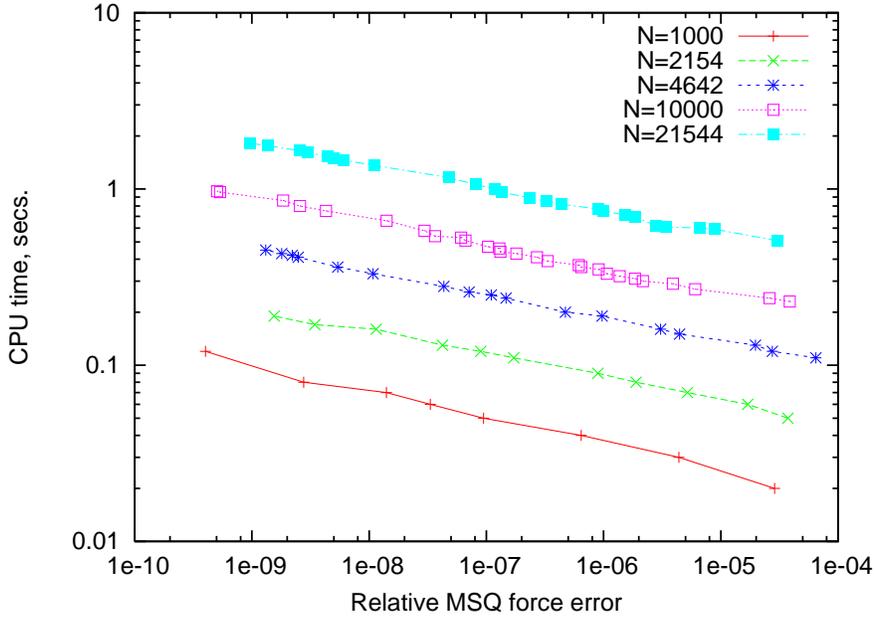}
\caption{The Pareto frontiers for the random system on Opteron 244. }\label{fig:pareto_random}
\end{figure}

\begin{figure}[htp]
\centering
\includegraphics[totalheight=.4\textheight,]{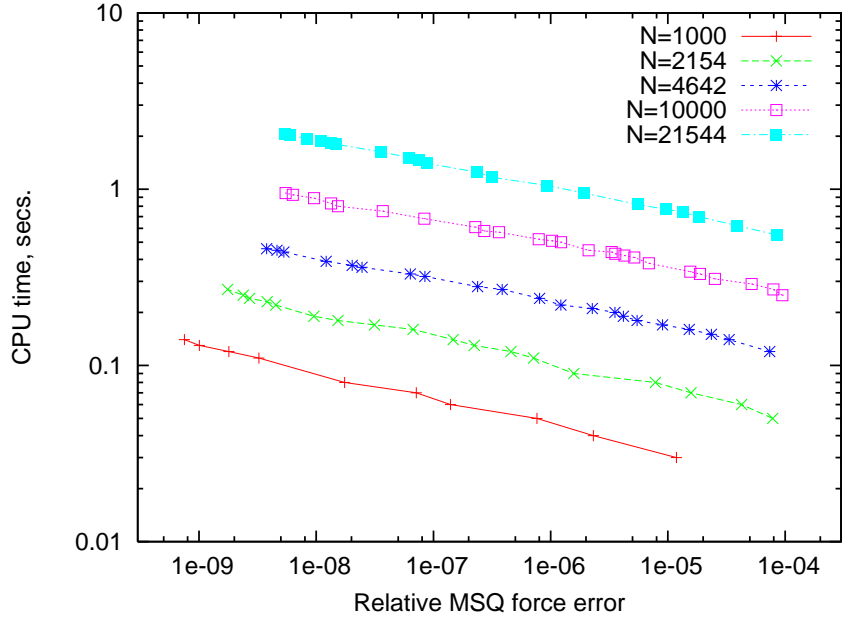}
\caption{The Pareto frontiers for the crystal system.}\label{fig:pareto_crystal}
\end{figure}

The Pareto frontiers for the values of $N=1000\times 10^{j/3}$; $j=0\dots 4$ for the random
system and the crystal systems are given at the Fig.\ref{fig:pareto_random} and Fig.\ref{fig:pareto_crystal}. The tests where done on an 1.8GHz AMD Opteron 244.

For the purpose of illustration we present here the optimal parameter values for the crystal system. The values for the random one do not differ significantly.

\begin{figure}[htp]
\centering
\includegraphics[totalheight=.4\textheight,]{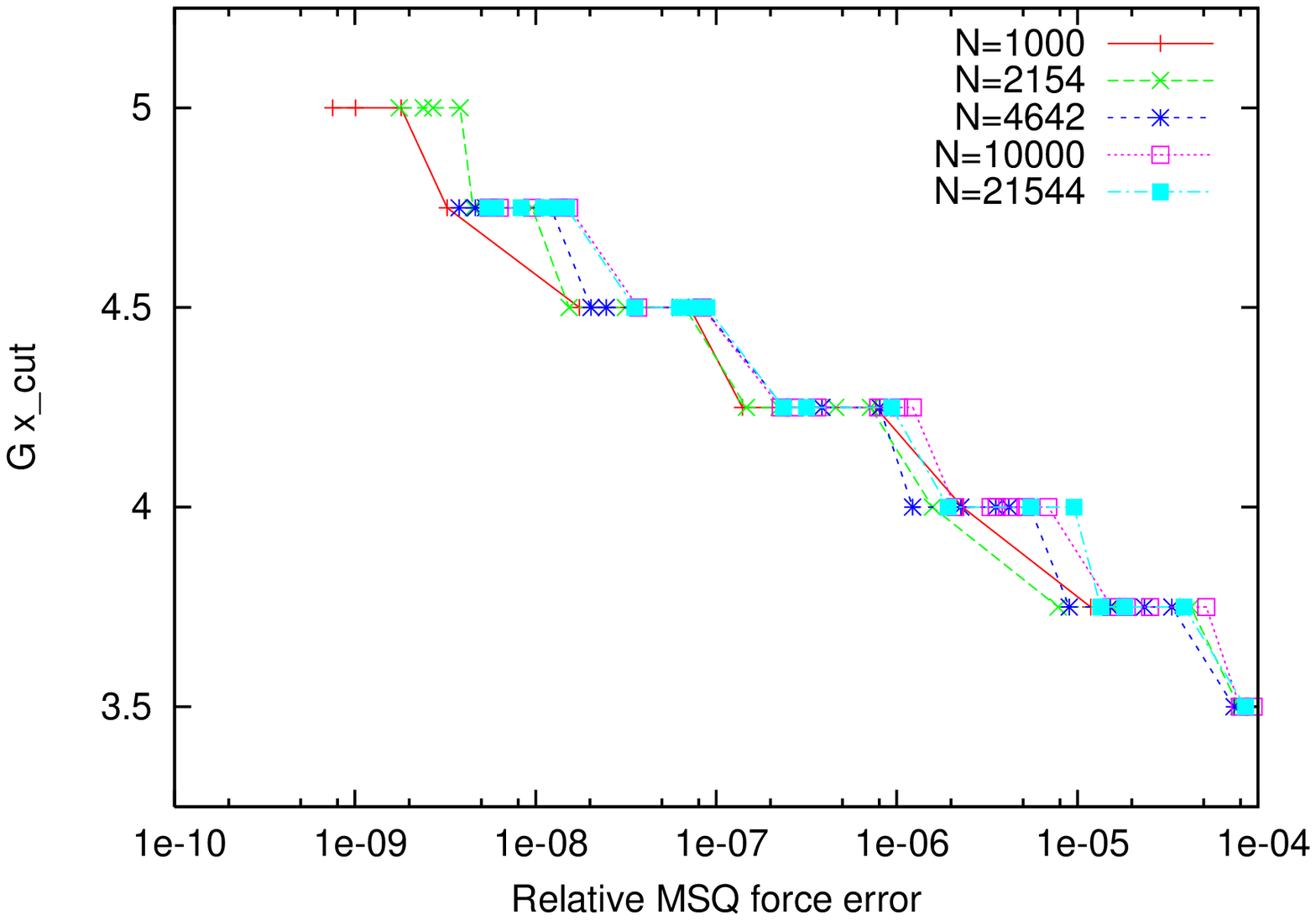}
\caption{The optimal values of $G\xcut$, crystal system.}\label{fig:xcut_crystal}
\end{figure}

\begin{figure}[htp]
\centering
\includegraphics[totalheight=.4\textheight,]{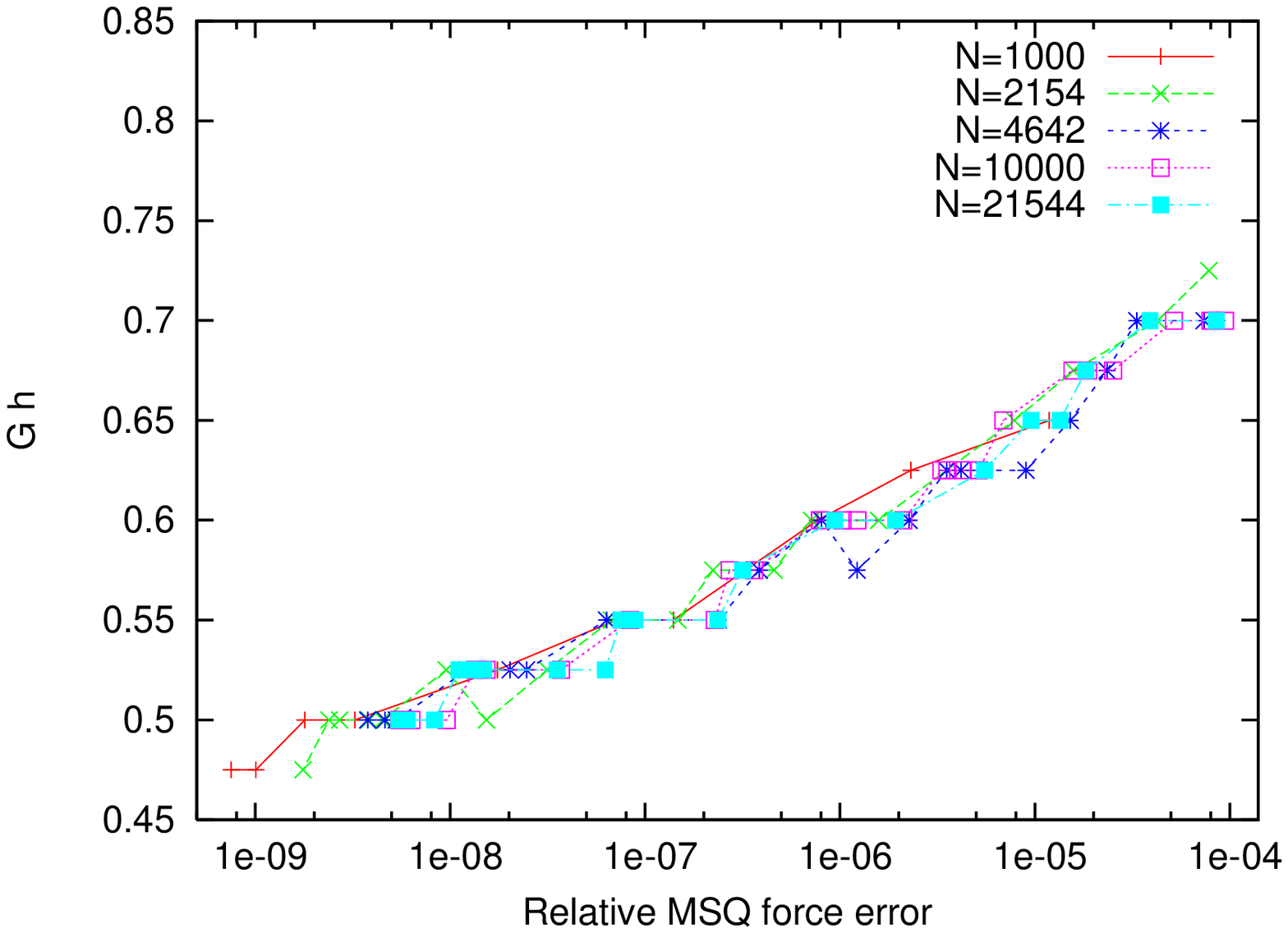}
\caption{The optimal values of $G h$, crystal system.}\label{fig:hgrid_crystal}
\end{figure}

\begin{figure}[htp]
\centering
\includegraphics[totalheight=.4\textheight,]{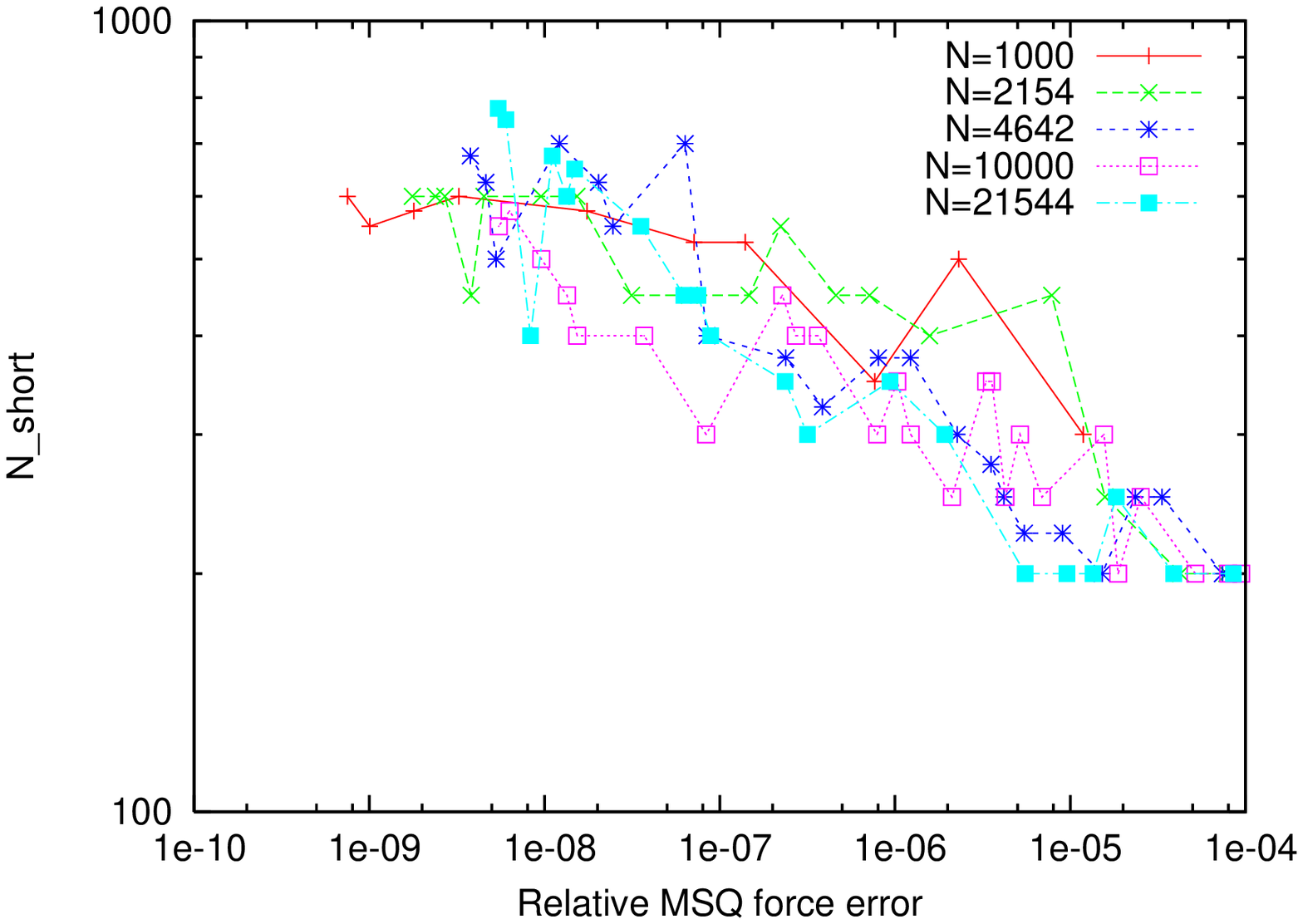}
\caption{The optimal values of $\nshort$, crystal system.}\label{fig:nshort_crystal}
\end{figure}

\begin{figure}[htp]
\centering
\includegraphics[totalheight=.4\textheight,]{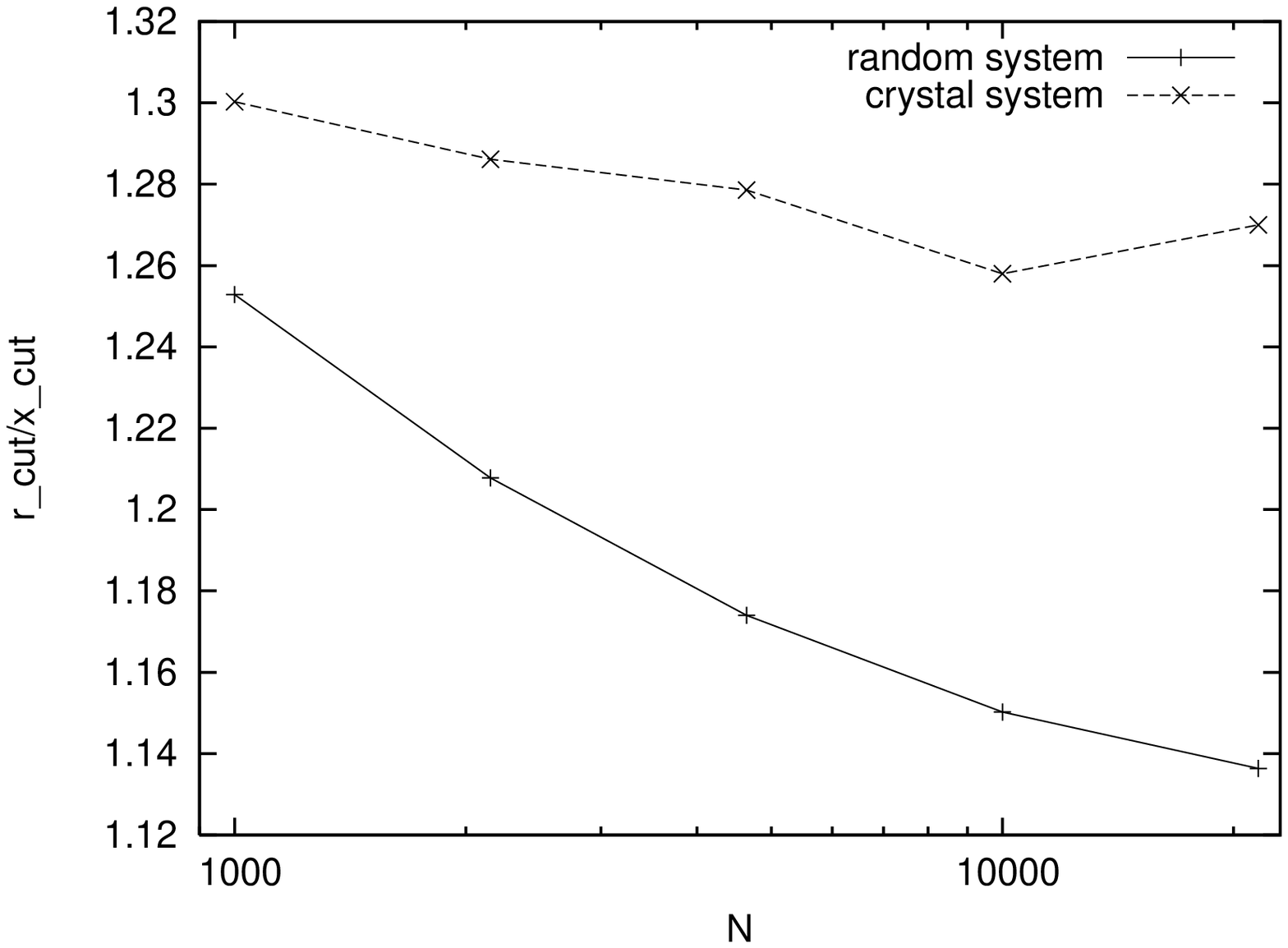}
\caption{Values of $\rcut/\xcut$ for both random and crystal systems.}\label{fig:factor}
\end{figure}

From Figs. \ref{fig:xcut_crystal},\ref{fig:hgrid_crystal} we see that the optimal values of $G\xcut$ and $G h$ are determined by the accuracy level. In contrast, the optimal value of  $N_{\rm short}$ depends on the number of particles too. Therefore in an actual calculation it must be adjusted by the trial and error method to give optimal accuracy and CPU time.
This is similar to finding the optimal value of the Gaussian width in the standard particle-mesh schemes \cite{hunen}.
Finally, the optimal ratio $\rcut/\xcut$ was observed to be independent on the required
accuracy but slightly decreased with rising $N$, as seen on Fig. \ref{fig:factor}.

\begin{figure}[htp]
\centering
\includegraphics[totalheight=.4\textheight,]{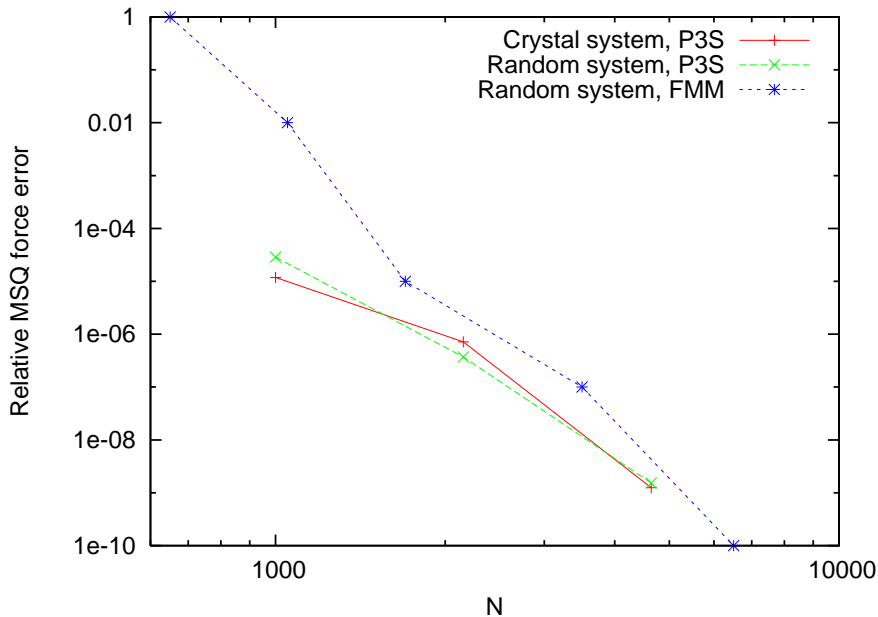}
\caption{The crossover  curves.}\label{fig:cross_crystal}
\end{figure}

The Pareto frontiers allow us to determine the crossover points for each $N$: the values of 
MSQ force error for which our calculation takes the same time as the full direct one. They
are presented at Fig.\ref{fig:cross_crystal}. We have plotted the crossover curves for the random system and for the crystal system. For comparison we have also plotted the crossover curve of the fast multipole method taken from \cite{Headgordon}, where one of the best  implementations of FMM is described. In that paper the same charge distribution was used as in our random system. Of course, the random positions of the particles where different in our tests and in those of \cite{Headgordon}. However, we still see that our method has  lower crossovers than the FMM for the same accuracy. 

\section{The energy conservation.}

In contrast to other methods such as FMM \cite{md}, \cite{schulten1}, our algorithm has the advantage that the approximate forces are exact analytic derivatives
of the approximate energy. This allows for energy conservation during an MD run. To illustrate this, we make an MD simulation of a rock salt crystal formed by 1000 Na and Cl atoms. The particle positions and velocities are updated by the velocity Verlet algorithm. 

To get physically reasonable results, we made the particles
interact through the Born-Mayer-Huggins-Fumi-Tosi (BMHFT) rigid-ion potential \cite{fumi} that has bonding terms in addition to the Coulombic force.

At first we made the system equilibrate for 300 oscillation periods. We then monitored the
potential and total energy for another 100 periods using the full direct algorithm. Then the last 100 periods were repeated using our P3S algorithm.

\begin{figure}[htp]
\centering
\includegraphics[totalheight=.4\textheight,]{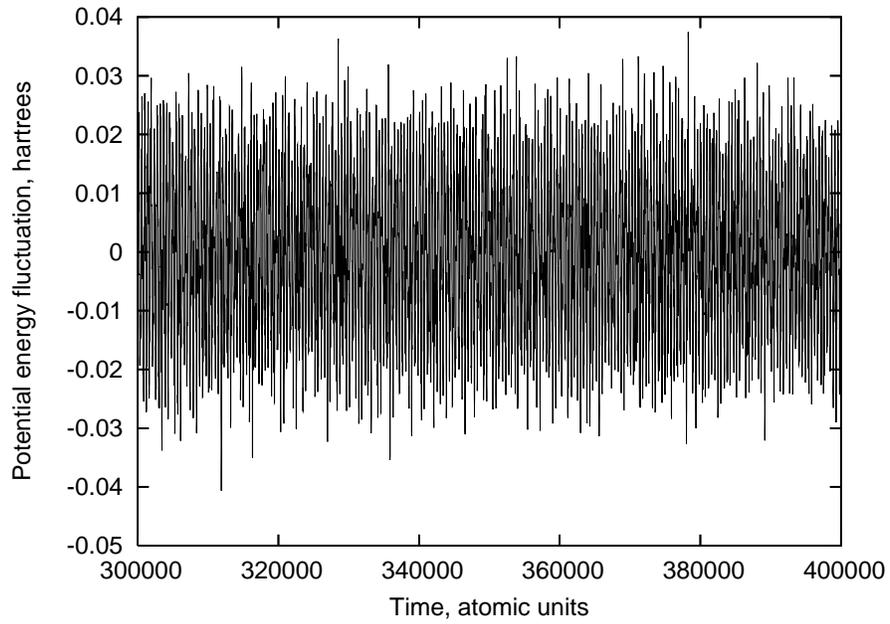}
\caption{The potential energy fluctuations with the P3S method.}\label{fig:epot}
\end{figure}

\begin{figure}[htp]
\centering
\includegraphics[totalheight=.4\textheight,]{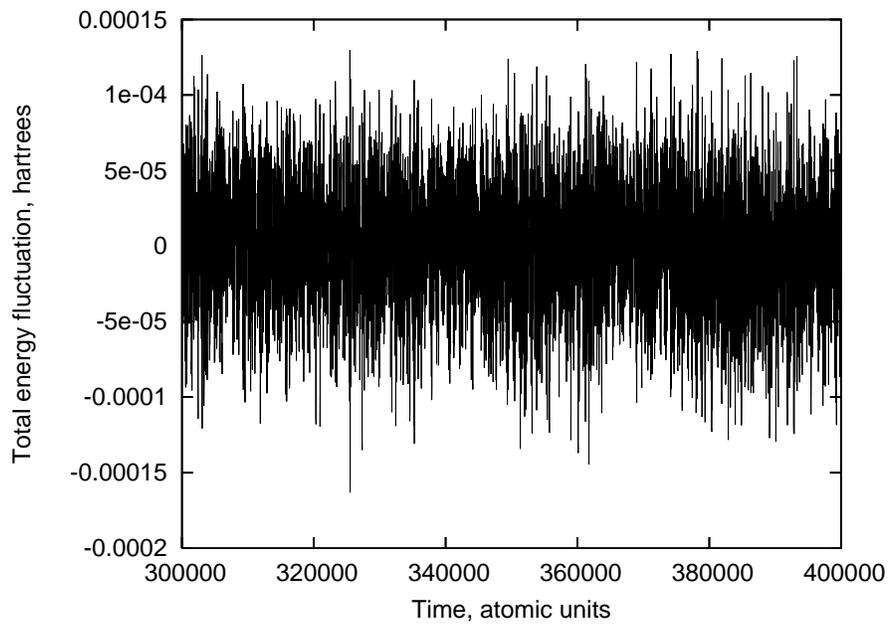}
\caption{The total energy fluctuations with the P3S method.}\label{fig:etot}
\end{figure}

On Fig. \ref{fig:epot}, \ref{fig:etot} we plot the absolute values of deviations of the potential and total energy from their
mean values. The ratio of the
mean square deviation of the total energy to that of the potential one is found to be equal to $1.4\times 10^{-3}$.

The P3S results are shown on the graphs; those of the full direct calculation are indistinguishable from the P3S ones.

\section{Parallelization.}

The parallelization of the calculation of the short-range energy and forces  is straightforward: for the particle $i$ and processor IPROC, we calculate the force  only if ${\rm MODULO(I,NPROC)=IPROC}$, where NPROC is the total number
of processors. In the end, an $\allreduce$ command sums all the contributions.

For the long-range part, we rely on the parallel structure
of the Poisson solver \cite{psolver}. The charge density on the grid is divided into slabs  in the $x-y$ plane. Each such slab is the input density at a separate node. The Poisson equation (for the whole cell) is solved for that slab density. The output at a given node again 
contains only the corresponding slab of the grid, this time it is a piece of
the potential array. Of course one needs global interprocessor communications, including ${\rm MPI\_ALLTOALL}$ for the solution of the Poisson equation.

Then in the original algorithm of \cite{psolver}, an $\allreduce$ command sums up the potentials of all the slabs.

In our program for point particles, we kept the parallelization of the
charge assignment as above: each processor receives only the grid charges from
the corresponding slab. However, we do not use the $\allreduce$ command
to get the potential. Instead, we used the slabwise structure of the output of the Poisson solver. For each node, the corresponding slab potential contributes only to the forces on particles that are close to it. 

In the end, we add up the ${\it forces}$ with the $\allreduce$ command.
This $\allreduce$ is actually merged in the program with the one for 
the short-range forces. 

In this way we minimize the interprocessor communication considerably compared
e.g. to \cite{pollock} since it is much easier to send the components of the
forces than the pieces of the enormous potential array.

To test the parallel program we ran it for the random system (the same as in the serial tests) with $10^5$ particles. The relative accuracy of the forces was kept around $10^{-6}$. 

\begin{figure}[htp]
\centering
\includegraphics[totalheight=.4\textheight,]{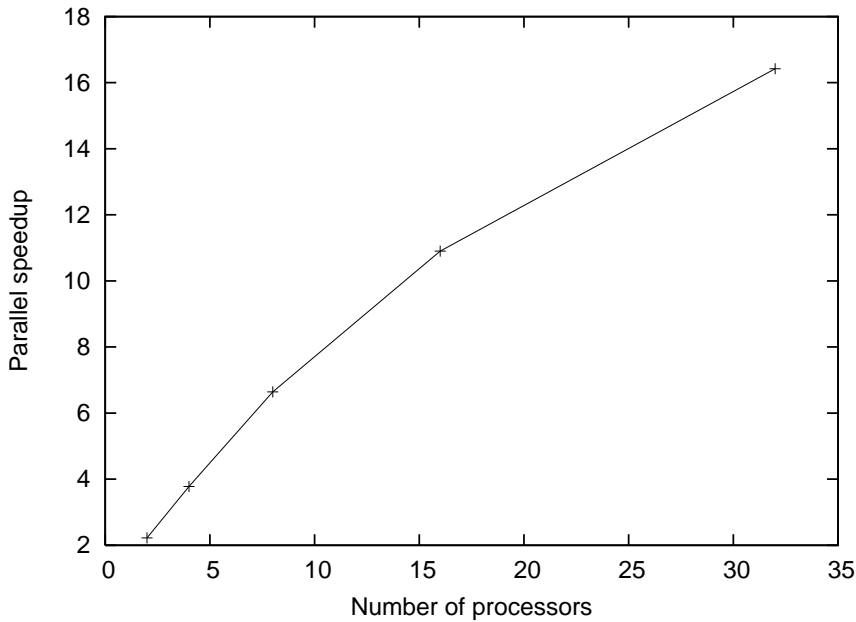}
\caption{The parallel speedup results on a CRAY XT3 for $10^5$ particles.}\label{fig:speedup}
\end{figure}

The result is given at Fig.\ref{fig:speedup}. At the $y$ axis we have the ratio of the 
CPU time spent on several processors to that spent by one processor. 


\section{Conclusion.}

We have developed a point particle Poisson solver algorithm that has a lower crossover point than the FMM. It can be considered as a generalization of the particle-mesh solvers for free boundary conditions. It can also achieve high precision.

The forces obtained by our program are
analytical derivatives of the energy; this is an advantage in the context of MD simulations.
An MPI parallelised version of the algorithm is presented that scales well on a moderate number of processors.

\section{Acknowledgments.}

The authors acknowledge financial support from the Swiss National Science Foundation. The numerical calculations were done on a parallel supercomputer of the Swiss National Supercomputing Center.

\end{document}